\documentclass[conference]{IEEEtran}
\IEEEoverridecommandlockouts
\usepackage{cite}
\usepackage{amsmath,amssymb,amsfonts}

\usepackage{graphicx}
\usepackage{textcomp}
\usepackage{balance}
\usepackage[all]{nowidow}
\usepackage{xcolor}
\usepackage{subcaption}
\usepackage{bookmark}
\usepackage{tabularx}
\usepackage{booktabs}
\usepackage{array}
\usepackage{amsmath}

\usepackage{algorithm}
\usepackage[noend]{algpseudocode}
\usepackage{caption}
\usepackage{listings}
\usepackage{siunitx}
\usepackage{fnpct}

\usepackage[capitalise]{cleveref}

\usepackage{xparse}
\newcommand{\name}{Konflux}
\begin{document}

\title{\textsc{Konflux}: Optimized Function Fusion for Serverless Applications}
\author{\IEEEauthorblockN{Niklas Kowallik, Trever Schirmer, David Bermbach}
    \IEEEauthorblockA{\textit{Technische Universit\"at Berlin}\\
        \textit{Scalable Software Systems Research Group} \\
        \{nk,ts,db\}@3s.tu-berlin.de}
}

\maketitle

\begin{abstract}
    Serverless platforms enable fine-grained deployment of \emph{Function-as-a-Service} (FaaS) applications, yet optimizing function fusion remains challenging due to the combinatorial explosion of possible configurations.
    Even moderately sized applications induce configuration spaces that render exhaustive cloud benchmarking prohibitively expensive.
    We present \name{}, a local FaaS emulation framework for rank-order optimization of fusion configurations.
    \name{} executes applications locally while integrating a provider-specific cost model, enabling systematic exploration of the full configuration space without repeated cloud deployment.
    The approach serves as a middle ground between cost-intensive cloud benchmarking and lightweight heuristics or simulation, balancing experimentation cost and modeling fidelity.
    Across multiple applications and resource limits, we evaluate all possible fusion configurations.
    We show that only a small fraction of configurations performs near-optimal, and that pricing models significantly shape the optimal fusion strategy.
    Validation on AWS Lambda demonstrates that configurations ranked highly by \name{} fall within the 5\% top-performing sets in the cloud, confirming that the framework reliably captures qualitative optimization trends at a fraction of the cost of exhaustive cloud benchmarking.
    \name{} is available as open source.\footnote{\href{https://github.com/nkowallik/konflux}{https://github.com/nkowallik/konflux}}
\end{abstract}


\section{Introduction}
\label{sec:introduction}

\emph{Function-as-a-Service} (FaaS) has emerged as a widely adopted cloud service model, offering scalability and operational simplicity~\cite{foxStatusServerlessComputing2017,eismannwhy,cvetkovicDirigentLightweightServerless2024,paper_pfandzelter2020_tinyfaas}. 
In FaaS, applications are decomposed into fine-grained functions that are fully managed by the provider~\cite{akhtarCOSEConfiguringServerless2020a,paper_bermbach2020_faas_coldstarts,cordinglyFunctionMemoryOptimization2022}. 
Despite abstracting infrastructure management, FaaS introduces new challenges in latency optimization, cost control, and orchestration~\cite{wangPeekingCurtainsServerless2018,aslanpourServerlessEdgeComputing2021}. 
Network communication between functions contributes significantly to overall cost, as inter-function traffic is billed and increases latency~\cite{elgamalCostlessOptimizingCost2018,klimovicPocketElasticEphemeral2018}. 
Synchronous invocations further exacerbate this issue: While a function waits for a callee to return, it continues to incur billing, leading to so-called double billing~\cite{baldiniServerlessTrilemmaFunction2017a}. 

Function fusion has been proposed as a mechanism to mitigate these effects by combining multiple fine-grained functions into a single deployment unit~\cite{elgamalCostlessOptimizingCost2018,tirkeyNovelFunctionFusion2023,schirmer2024fusionizepp,liEnhancingEffectiveBidirectional2024}. 
By redeploying applications with fused functions and adapted resource limits, prior work demonstrates measurable cost and latency improvements~\cite{paper_schirmer2022_fusionize,schirmer2024fusionizepp}. 
Following prior work~\cite{schirmer2024fusionizepp}, we refer to such combined deployment units as \emph{fusion groups}. 

\begin{figure}
    \centering
    \includegraphics[width=\linewidth]{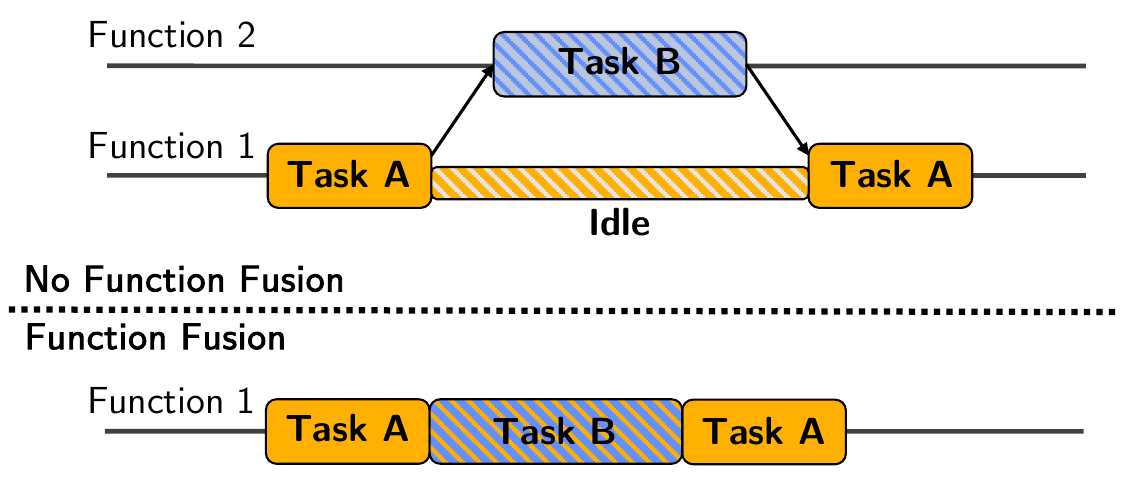}
    \caption{Without Function Fusion the synchronous call of \texttt{Task A} to \texttt{Task B} results in an idle time for \texttt{Function 1}, which is the sum of the execution duration, network latency, and warm/cold function start latency of the synchronously called \texttt{Task B}.
    With Function Fusion that idle time is used for the local execution of \texttt{Task B}, eliminating the need for an additional instance and the network overhead.}
    \label{fig:function_fusion_motivation}
\end{figure}

However, identifying the optimal fusion configuration for a given application remains challenging. 
The configuration space grows combinatorially with the number of functions and resource limits. 
Heuristic and analytical approaches simplify this space through assumptions about platform behavior or workload characteristics~\cite{elgamalCostlessOptimizingCost2018,tirkeyNovelFunctionFusion2023}. 
While efficient, these approaches trade accuracy for tractability. 
At the other extreme, exhaustive cloud benchmarking represents the gold standard for obtaining exact cost and latency values, but is prohibitively expensive for realistic applications due to the size of the parameter space.

This work addresses the gap between lightweight heuristics and costly cloud benchmarking. 
We propose \name{}, a local FaaS emulation framework for systematic rank-order optimization of fusion configurations. 
\name{} executes applications locally while incorporating a provider-specific cost model, enabling exhaustive exploration of fusion configurations without repeated cloud deployment.
Emulating the target software yields higher functional fidelity than simulating it, as it executes the actual software stack rather than relying on behavioral models or heuristics.
The framework serves as a middle-ground solution: It provides higher fidelity than heuristic or simulation-based approaches, while reducing experimentation cost and turnaround time compared to full cloud benchmarking. 
This way, developers may validate the top-$k$ ranked configurations in the cloud, thereby limiting cloud experimentation to a small, targeted subset of candidates.

Using \name{}, we analyze the complete fusion configuration space of several applications under varying resource limits. 
Our results show that only a small fraction of configurations is near-optimal and that pricing models significantly influence the optimal fusion strategy. 
Moreover, we derive cross-application heuristics grounded in exhaustive emulation and validate their effectiveness against cloud-based measurements.

In summary, we make the following contributions:

\begin{enumerate}
    \item We present \name{}, a framework for exhaustive fusion configuration exploration via local FaaS emulation with an extensible cost model, and release it as open source (\cref{sec:systemdesign}).
    \item We perform a comprehensive evaluation across four applications and multiple resource limits, covering the full configuration space (\cref{sec:approach}).
    \item We demonstrate that only a limited subset of configurations is near-optimal, validate rank-order predictions against cloud experiments, and derive empirically grounded fusion heuristics that extend prior work (\cref{sec:discussion}).
\end{enumerate}
\section{Background \& Related Work}
\label{sec:related_work}
\label{sec:background}

FaaS introduces a cloud execution model in which applications are decomposed into discrete functions that are managed entirely by the platform~\cite{Jonas_2019_ViewOn}. 
The platform handles scaling, scheduling, and instance management, while billing follows a pay-per-use model with fine-grained time accounting~\cite{lin_demystifying_2026}.

When no idle instance is available to serve an incoming request, the platform initializes a new execution environment, resulting in a \emph{cold start}.
Cold starts incur higher latency than warm invocations because the isolation environment and runtime have to be initialized. 
In applications comprising multiple functions, these delays accumulate and significantly affect end-to-end latency.

Prior work has proposed multiple approaches to mitigate cold starts. 
Platform-level optimizations include startup CPU boosts (e.g., as seen in Google Cloud Run Functions) and alternative isolation mechanisms~\cite{Sahraei_2023_XFaaS}. 
Other techniques reduce initialization overhead~\cite{Brooker_2023_OnDemand}, rely on snapshotting~\cite{Brooker_2021_Uniqueness}, batch incoming requests~\cite{schirmer2023profaastinate,Sahraei_2023_XFaaS}, or maintain pre-warmed instances~\cite{216031}. 

Beyond cold starts, FaaS introduces additional performance and cost challenges.
Inter-function communication requires network traversal through the platform scheduler, increasing both latency and cost. 
Synchronous invocations are particularly expensive: While the requested function performs computation, the requesting function remains active and billed, leading to so-called double billing~\cite{baldiniServerlessTrilemmaFunction2017a}.

Function fusion addresses these inefficiencies by altering deployment granularity without modifying application logic. 
Developers express applications as logical tasks; traditionally, each task is deployed as an individual function. 
Fusion frameworks instead deploy multiple tasks jointly as a single function, referred to as a \emph{fusion group}~\cite{paper_schirmer2022_fusionize,schirmer2024fusionizepp}. 
By eliminating network communication between synchronously invoked tasks, fusion reduces both latency and double billing (Figure~\ref{fig:function_fusion_motivation}). 
Some systems further adapt resource configurations dynamically to react to workload changes~\cite{schirmer2023nightshift}.

Identifying optimal fusion groups remains a central challenge. 
The configuration space grows combinatorially with the number of tasks and resource limits. 
Several approaches model cost and performance analytically to guide search. 
Lin et al.~\cite{linFineGrainedPerformanceCost2023} employ conditional stochastic nets to model cost and latency, enabling programmatic search over fusion and bundling decisions. 
Costless~\cite{elgamalCostlessOptimizingCost2018} formulates joint placement and fusion as a linear optimization problem using a cost graph abstraction. 
Wisefuse~\cite{mahgoubWISEFUSEWorkloadCharacterization2022} leverages DAG-based workload characterization to guide bundling decisions. 
Other works rely on DAG abstractions combined with heuristic pruning strategies~\cite{mahgoub_orion_2022,linFineGrainedPerformanceCost2023,schirmer2024fusionizepp,wenJointOptimizationParallelism2024,rameshOptimalMappingWorkflows2024}. 
For example, Schirmer et al.~\cite{schirmer2024fusionizepp} propose fusing synchronously invoked tasks while separating asynchronous ones.

While these approaches can significantly reduce search complexity, they typically rely on simplifying assumptions or more constrained exploration strategies.
As a result, there may be scenarios—especially under realistic platform dynamics and pricing models—where the global optimum across the full fusion configuration space is not fully explored.
\section{\name{}: Design \& Implementation}
\label{sec:systemdesign}

To support systematic exploration of the fusion configuration space, a platform has to provide (i) fine-grained observability of function executions, (ii) control over invocation and communication paths, and (iii) cost-efficient experimentation independent of public cloud infrastructure.
In particular, exhaustive evaluation requires precise measurement of execution behavior while avoiding the prohibitive cost of repeated cloud deployment.

\begin{figure}
    \centering
    \includegraphics[width=\linewidth]{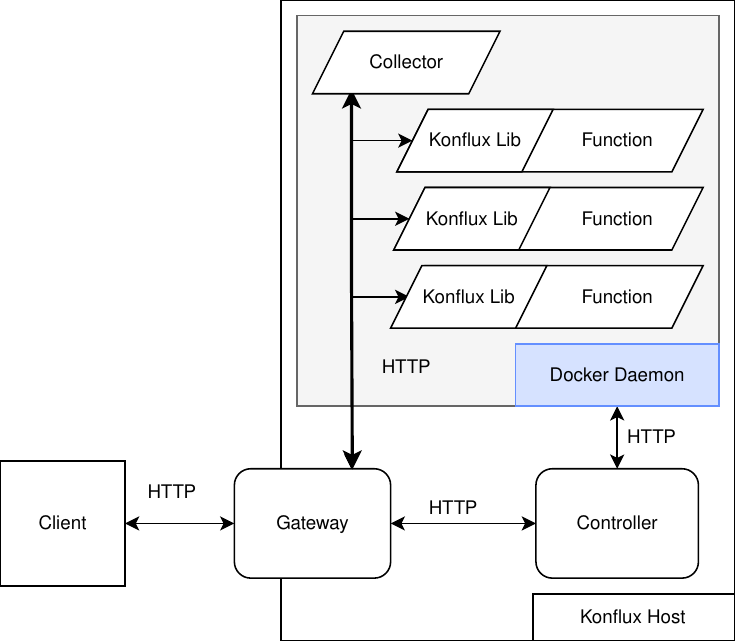}
    \caption{\name{} architecture overview.
    The client communicates with a gateway, which forwards requests to functions running inside Docker.
    The gateway relays control messages to a controller, which starts and stops Docker instances.
    A centralized collector collects metrics from all components.}
    \label{fig:conflux}
\end{figure}

We introduce \name{}, a framework for exhaustive exploration of fusion-group configurations via local FaaS emulation.
\name{} executes applications in a controlled environment that reproduces relevant execution semantics of FaaS platforms while integrating a provider-specific cost model.
This design enables rank-order optimization across the complete configuration space without requiring cloud-based experimentation for every candidate.

\name{} follows a middle-ground philosophy between pure simulation frameworks, heuristic-based estimations, and direct benchmarking on commercial cloud providers.
It executes real application code locally to capture realistic runtime and communication behavior, while abstracting provider-specific billing through an extensible cost model.
A central design requirement of \name{} is the integration of a cost model that enables the evaluation of fusion configurations not only with respect to runtime performance, but also in terms of their monetary impact.

By performing experiments locally, \name{} allows exhaustive benchmarking of all fusion setups at substantially lower cost and turnaround time compared to cloud experimentation.
Developers can subsequently validate the top-$k$ ranked configurations in the cloud, reducing public cloud deployment to a targeted verification step rather than an exploratory search process.

\name{} is implemented in Go, with a lightweight Python-based function library component, and uses Docker containers as the execution substrate to ensure portability and reproducibility.
The framework is open-source\footnote{\href{https://github.com/nkowallik/konflux}{https://github.com/nkowallik/konflux}} and designed for modularity and extensibility.
Its architecture consists of three core components: the \emph{gateway}, the \emph{controller}, and the \emph{collector} (Figure~\ref{fig:conflux}).

\textbf{Gateway.} The gateway acts as a reverse proxy that receives external requests and forwards them to the appropriate function containers.
It mediates client–function communication and intercepts invocation metadata required for measurement.

\textbf{Controller.} The controller manages the lifecycle of function containers, including deployment, teardown, and cleanup.
It enforces fusion configurations by orchestrating container composition and resource limits for each experiment.
Each fusion group is deployed as its own container with its own resource limit, mirroring the one-instance-per-deployment-unit model of commercial FaaS platforms; a configuration of $k$ groups therefore runs as $k$ containers, not as a single one.
Tasks within a group communicate through in-process calls, whereas calls that cross a group boundary traverse the gateway and are measured as remote invocations.
Fusion is thus expressed as a change in container topology, which is precisely the effect being studied.

\textbf{Collector.} The collector aggregates fine-grained execution metrics for each request, including runtime characteristics and inter-function communication events.
These measurements form the basis for cost computation and ranking.

Together, these components provide a local FaaS execution environment that enables exhaustive benchmarking of fusion configurations under realistic execution semantics.
By combining measured execution behavior with a configurable billing model, \name{} produces emulation-based rankings of fusion setups while avoiding the cost-prohibitive nature of full cloud exploration.

\subsubsection{Costs \& Sorting}
\label{sec:systemdesign:costs}

A key component of \name{}’s design is the integration of a cost model to evaluate fusion configurations.
The system combines execution metrics such as runtime and end-to-end latency  with configurable cost parameters to produce a score for each configuration:

\begin{equation} \label{eq:1}
    score = \alpha \cdot latency + (1-\alpha) \cdot cost
\end{equation}

The parameter $\alpha \in [0,1]$ defines a continuous trade-off between latency and cost, effectively encoding user preference into the objective function.
When $\alpha = 1$, the score depends solely on latency (latency minimization), whereas $\alpha = 0$ reduces the objective to cost minimization.
Intermediate values of $\alpha$ yield weighted compromises between both metrics.

This formulation induces a family of objective functions parameterized by $\alpha$, allowing systematic exploration of different optimization preferences without modifying the underlying model.
By varying $\alpha$, configurations can be evaluated and rank-ordered according to arbitrary user-defined priorities between latency and cost.

Consequently, this approach enables consistent comparison across configurations and supports structured analysis of the trade-off surface between latency and cost.

Decoupling execution measurement from cost evaluation allows the design to allow for new cost models or scoring criteria, which can be introduced without modifying the execution layer.
This modular approach allows the framework to adapt to different evaluation policies or provider-specific billing schemes, while maintaining exhaustive and efficient analysis of fusion setups.

\section{Experiment Design}
\label{sec:approach}
Our goal is rank-order optimization of fusion setups rather than exact value prediction. 
Accordingly, we enumerate the complete configuration space and analyze relative performance under different trade-off regimes.

Applications are represented as directed acyclic graphs (DAGs), where nodes denote tasks and edges denote invocation relationships. 
Fusion groups are constructed along DAG edges, allowing adjacent tasks to be combined into a single deployment unit. 
This representation follows established practice in fusion research~\cite{schirmer2024fusionizepp,czentyeServerlessApplicationComposition2024,mahgoubWISEFUSEWorkloadCharacterization2022,sheshadriRightFusionEnablingQoS2024}.

Although the configuration space grows combinatorially with the number of tasks and resource limits, a central hypothesis of this study is that only a small subset of configurations exhibits practically meaningful trade-offs. 
The exhaustive exploration enabled by \name{} allows us to empirically identify these configurations instead of relying on pruning heuristics or analytical assumptions.

We benchmark \emph{every} fusion-group configuration under \emph{all} resource limits shown in \cref{tab:limits}. 
The selected configurations reflect the proportional CPU-to-memory scaling model of AWS Lambda.
While AWS also scales I/O and network throughput with memory, our emulation focuses on computational behavior and invocation overhead.
The enumerated space thus combines every fusion decision along the DAG edges with every assignment of the three resource levels to the resulting groups.
For our applications this yields 768 configurations for the five-task topologies and 12{,}288 for the seven-task \texttt{TREE}.
The space grows combinatorially with the number of tasks, which bounds what exhaustive exploration can cover in practice (\cref{sec:discussion}).

\begin{table}
    \centering
    \begin{tabularx}{.4\linewidth}{@{}>{\centering\arraybackslash}X>{\centering\arraybackslash}X>{\hspace*{2ex}\raggedright\arraybackslash}Xl@{}}
        \toprule
        \bf{CPU} & \multicolumn{1}{c}{\bf{RAM}} \\
        \midrule
        0.1 & \SI{128}{MB}\\
        0.5 & \SI{832}{MB}\\
        1.0 & \SI{1769}{MB}\\
        \bottomrule
    \end{tabularx}
    \caption{Resource limits used for the function and fusion group instances in benchmarking, reflecting AWS Lambda~\cite{awsLambda} resource scaling for fixed CPU associated values.}
    \label{tab:limits}
\end{table}

In summary, the experimental design is structured to (i) exhaustively characterize the fusion configuration space, (ii) enable workload-specific rank-order optimization, (iii) derive cross-application heuristics grounded in exhaustive evaluation, and (iv) validate transferability to production cloud environments.

\subsection{Benchmarking Applications}
\label{sec:approach:applications}
Our benchmark applications are inspired by prior work~\cite{schirmer2024fusionizepp}. 
The objective is to isolate the effects of fusion and resource allocation while maintaining controlled computational characteristics. 
All input parameters are fixed, and workloads use the \emph{Lucas--Lehmer test} to generate deterministic and sustained CPU load.
We select this kernel because it is work-bounded rather than time-bounded: fixing the Mersenne exponent $p$ fixes the iteration count and operand width, leaving duration as the dependent variable that fusion is expected to change.
General-purpose generators such as \texttt{stress-ng} are timeout-driven and execute as external processes rather than in-process application code, and can therefore neither be fused nor deployed unmodified on AWS Lambda.
The four topologies are chosen to vary the one property that fusion acts upon, namely invocation semantics, while holding the workload kernel constant: \texttt{LINEAR} isolates purely synchronous chains, \texttt{ASYNC} the purely asynchronous case, and \texttt{PARALLEL-LINEAR} and \texttt{TREE} the mixed regimes with increasing branching.
Any effect we attribute to fusion is thus a property of the invocation structure rather than of the computation being fused.

\begin{figure}
    \centering
    \includegraphics[width=.7\linewidth]{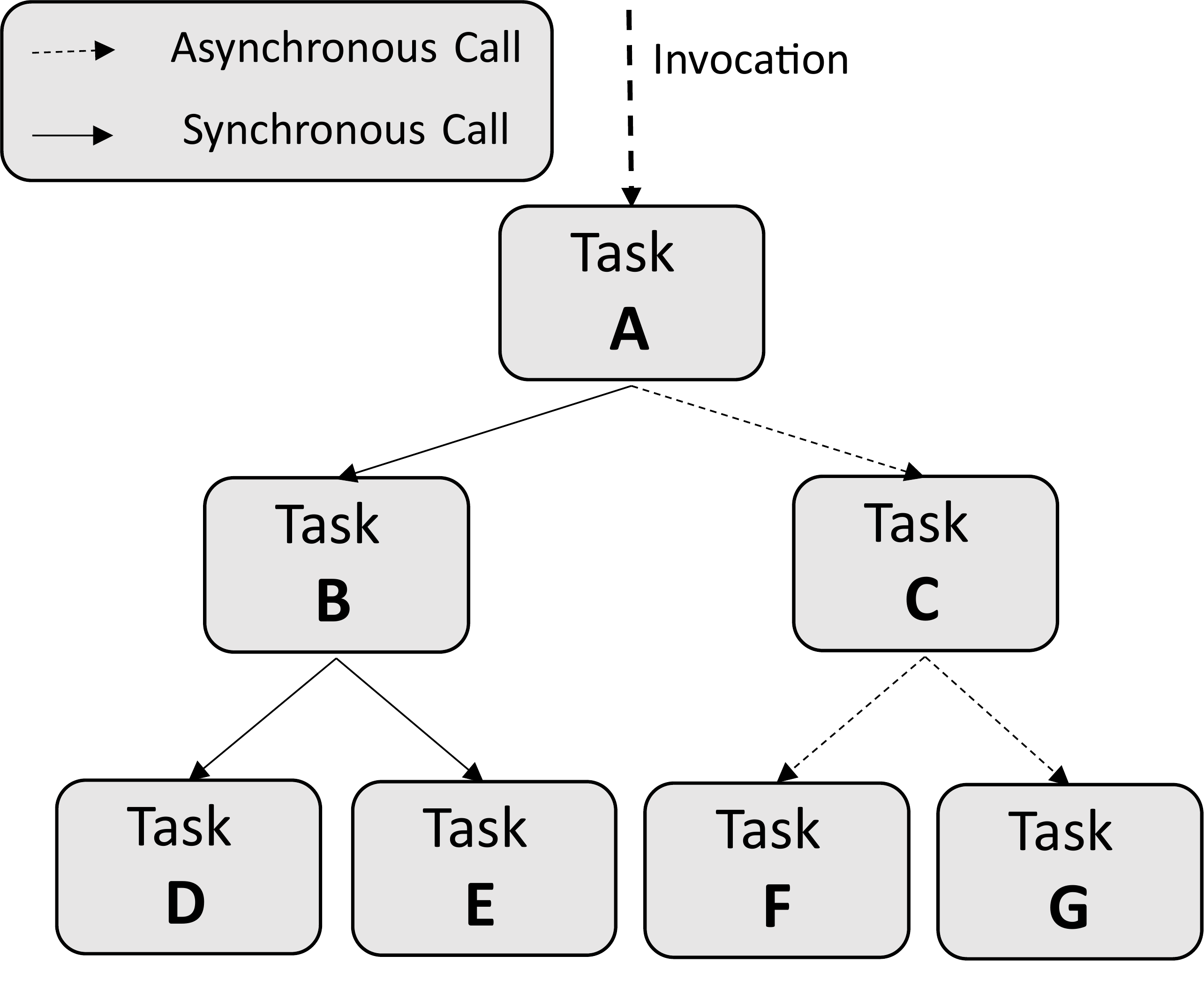}
    \caption{The \texttt{TREE} application splits after task \texttt{A} into a synchronous branch over tasks \texttt{B}, \texttt{D}, and \texttt{E}, and an asynchronous branch over tasks \texttt{C}, \texttt{F}, and \texttt{G}. Tasks \texttt{A}, \texttt{B}, \texttt{D}, and \texttt{E} are invoked synchronously with moderate computational load, whereas tasks \texttt{C}, \texttt{F}, and \texttt{G} are invoked asynchronously with heavy computational load.}
    \label{fig:tree}
\end{figure}

\textbf{TREE.}
The \texttt{TREE} application combines synchronous and asynchronous branches (Figure~\ref{fig:tree}). 
Task \texttt{A} invokes \texttt{B} synchronously and \texttt{C} asynchronously; \texttt{B} in turn calls \texttt{D} and \texttt{E} synchronously, while \texttt{C} calls \texttt{F} and \texttt{G} asynchronously.
Tasks \texttt{A}, \texttt{B}, \texttt{D}, and \texttt{E} thus form the synchronous branch with moderate load, whereas \texttt{C}, \texttt{F}, and \texttt{G} form the asynchronous branch with higher computational intensity.
This topology captures heterogeneous workloads and enables detailed analysis of fusion under mixed invocation semantics.

\textbf{LINEAR.}
Five tasks (\texttt{A-E}) invoke each other synchronously in sequence.
This topology isolates the effects of eliminating double billing through fusion of synchronous chains.

\begin{figure}
    \centering
    \includegraphics[width=0.4\linewidth]{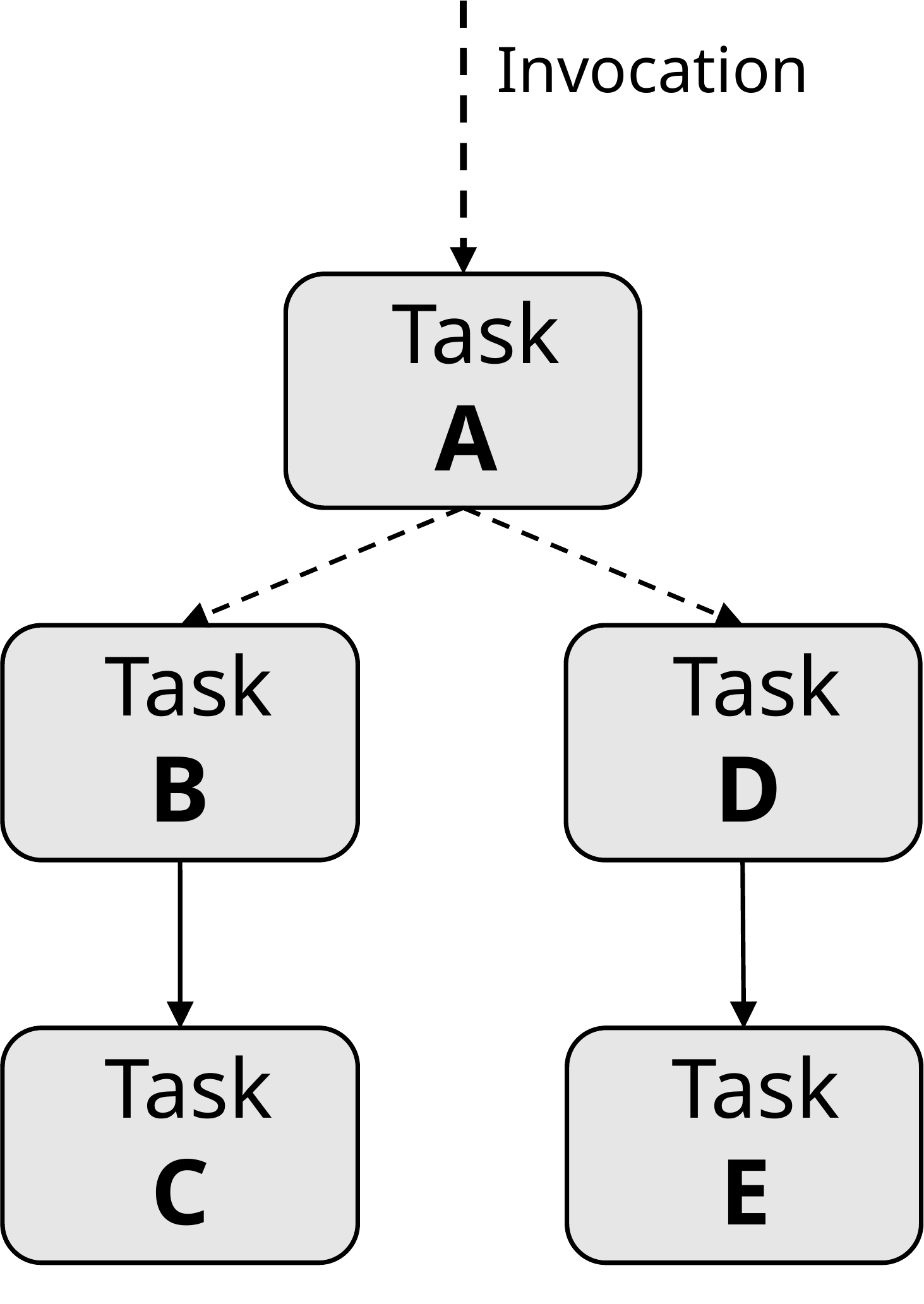}
    \caption{The \texttt{PARALLEL-LINEAR} application initiates parallel asynchronous calls to \texttt{B} and \texttt{D}, which themselves each have another asynchronous call with their respective following functions \texttt{C} and \texttt{E}.}
    \label{fig:parallel-linear}
\end{figure}

\textbf{PARALLEL-LINEAR.}
A variation of \texttt{LINEAR}, where tasks \texttt{B} and \texttt{D} initiate asynchronous branches (Figure~\ref{fig:parallel-linear}). 
This creates two parallel invocation paths and allows analysis of fusion under partial concurrency.

\textbf{ASYNC.}
A fully asynchronous variant of \texttt{LINEAR}. 
This topology isolates the effect of fusion when no synchronous waiting occurs

By evaluating all structural and resource combinations for each topology independently, \name{} enables fine-grained analysis of how fusion interacts with invocation semantics and workload characteristics. 
This exhaustive characterization forms the empirical basis for deriving generalizable heuristics.

\subsection{\name{} Study}

The benchmarking data produced by \name{} is combined with provider pricing models to compute invocation costs. 
We evaluate pricing schemes of AWS Lambda and Google Cloud Run Functions.

AWS Lambda charges a per-request fee and a duration-based fee per millisecond. 
Cloud Run Functions supports a similar model and an alternative runtime-only model. 
These models are selected to represent common production billing schemes and are not intended as a platform comparison.

The current prototype implements the Google Cloud Run Functions and AWS Lambda pricing model, but the architecture allows substitution of alternative provider models without changes to the execution layer.

\subsubsection{Costs \& Sorting}
\label{sec:approach:costs}
\label{sec:approach:sorting}
The detailed data generated by running benchmarks in \name{} allows us to calculate invocation costs according to different pricing models.
In this evaluation, we use the pricing models from the two most popular cloud providers AWS Lambda and Google Cloud Run Functions.
In Lambda, costs are based on a per-request fee and a fee per millisecond of runtime.
While Cloud Run Functions can also use this billing model, it also offers another model, where costs are only based on the runtime (and not per request).
Note that these pricing models are chosen to cover a wide variety of models that are used in production, and are not meant as a comparison of the platforms.
All costs are displayed in \$ per one million invocations (\${}pmi).

\subsubsection*{Scoring Setups}
For evaluation, \name{} applies the cost model to compute a weighted score for each fusion configuration (see \cref{eq:1}).

The parameter $\alpha \in [0,1]$ controls the trade-off between latency and cost.
We stress that $\alpha$ does not parameterize the fusion itself: the set of configurations is fixed by the application graph and the resource levels, and each configuration is measured exactly once.
Sweeping $\alpha$ only re-ranks these pre-measured configurations, so each value of $\alpha$ selects one winner from the same fixed set of candidates.
In our experiments, we implement AWS Lambda and Google Cloud Run Functions as two concrete pricing schemes, but the architecture allows substitution of alternative provider models without changes to the execution layer.
We evaluate 10{,}001 values of $\alpha$, defined as

\[
\alpha = \frac{\gamma}{10{,}000}, \quad \gamma \in \{0, \dots, 10{,}000\},
\]

where $\gamma$ is an integer index representing the discrete steps used to sweep the parameter space from pure cost optimization ($\alpha=0$) to pure latency optimization ($\alpha=1$). 
This produces a complete family of rank-orderings across all possible developer preferences.
The discretization into 10,001 values corresponds to a step size of 0.0001 (0.01 percentage points). This allows us to distinguish, for example, between weightings of 50.00\,\% and 50.01\,\%, while differences below this threshold (e.g., 50.009\,\% vs. 50.01\,\%) are intentionally treated as equivalent.
Because scoring operates on pre-collected execution and communication metrics, this step introduces negligible computational overhead.

\subsubsection{Setup}
\label{setup}
All experiments are conducted on a virtual machine running Ubuntu 22.04.5 LTS with 12 KVM cores at \SI{2095.076}{MHz}, \SI{16384}{KB} cache, and \SI{18432}{MiB} RAM. 
Requests are generated from a Raspberry Pi 5 with \SI{8192}{MiB} RAM. 

\subsection{Feasibility Study on AWS Lambda}

To assess whether emulation-based rankings transfer to real deployments, we perform a targeted validation on AWS Lambda.
While cloud benchmarking provides ground-truth latency and billing measurements, exhaustive evaluation of all fusion configurations in the cloud would be prohibitively expensive.

Instead, we select a representative subset of configurations that \name{} found, containing configurations from top, center and lower performance areas.
This focused evaluation tests whether the qualitative trends identified through emulation, such as fusing synchronous invocation chains while isolating asynchronous branches, hold under actual AWS runtime and billing semantics.

For each selected configuration, we measure end-to-end latency and monetary cost in AWS Lambda and compare the resulting rank order to the emulation-based predictions.
This analysis evaluates whether \name{} preserves rank-order behavior while significantly reducing the cost of exploratory benchmarking.

The fusion groups were deployed to \emph{europe-north-1} and their memory limits iterated, according to the limits we use in our experiments with \name{} (see \cref{tab:limits}).
\vfill\null
\section{Experiment Results}
\label{sec:evaluation}
This section presents a comprehensive evaluation of \name{} across diverse application topologies.
We first analyze the performance of our local emulation across 12,288 unique configurations to map the cost-latency trade-offs of function fusion.
This is followed by a feasibility study on AWS Lambda to validate the fidelity of our emulation against real-world cloud overheads.
Finally, we discuss key findings regarding optimal fusion patterns under both traditional and instance-based pricing models, demonstrating \name{}'s ability to provide reliable rank-order optimization.

\subsection{\name{} Study}
\begin{figure}
    \centering
    \includegraphics[width=\linewidth]{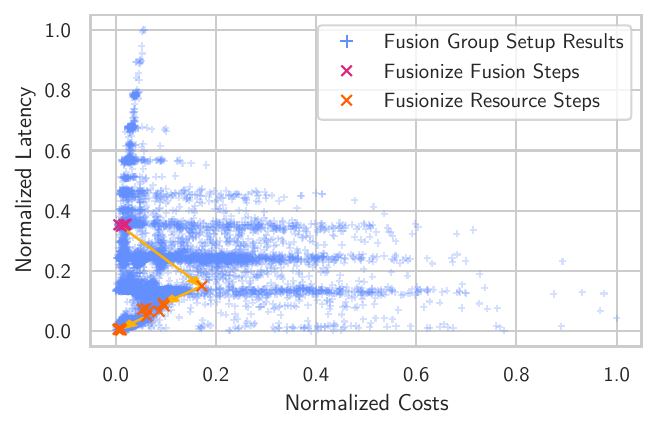}
    \caption{Benchmark results for the \texttt{TREE-FG} application shown as normalized latency versus normalized cost. The orange arrow traces the optimization trajectory of the state-of-the-art Fusionize approach~\cite{schirmer2024fusionizepp}, whose AWS Lambda--based decisions align with the optimal fusion group configurations identified by our emulation-based FaaS benchmarking platform. This illustrates how successive fusion and resource-allocation steps move the deployment toward the (near-)optimal region close to the origin relative to all possible fusion group setups. The traced path is an optimization trajectory, not a Pareto front; the Pareto-optimal set is the lower-left envelope of the point cloud, which the trajectory approaches but does not enumerate.}
    \label{fig:tree_scatter_fusionize}
\end{figure}

All costs displayed below are based on the AWS Lambda pricing model.
The results of the 12{,}288 unique fusion setups for the \texttt{TREE} application are displayed in \cref{fig:tree_scatter_fusionize}.
While the benchmarking results are spread across costs and latency, there is a large group of good performing fusion group setups near the graph's origin.
The graph also depicts an optimization path that displays a search for the best performing fusion group setup based on steps used in Schirmer et al.'s Fusionize approach~\cite{paper_schirmer2022_fusionize,schirmer2024fusionizepp}.
The steps of the optimization path are categorized as either changes to the composition of fusion groups (fusion step) or adjustments to the resource allocation within those groups (resource step).

\begin{figure}
    \centering
    \includegraphics[width=\linewidth]{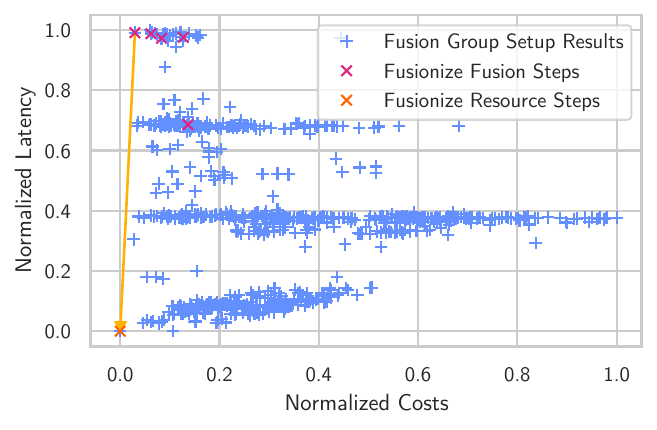}
    \caption{Benchmarking results for the \texttt{LINEAR} application shown as normalized latency versus normalized cost. 
    The highlighted optimization path of Fusionize~\cite{schirmer2024fusionizepp} mirrors the decisions derived from our emulation-based FaaS benchmarking platform, aligning with the optimal fusion group configurations across the explored design space. 
    As with \texttt{TREE-FG}, the iterative fusion and resource-allocation steps systematically drive the deployment toward the (near-)optimal region close to the origin relative to all possible fusion group setups.
}
    \label{fig:linear}
\end{figure}

The 768 unique benchmarking runs for the \texttt{LINEAR} application are displayed in \cref{fig:linear}.
There are 4 major result groups, which are close regarding latency, yet spread along the cost-axis.
The fusion of functions with the distinct resource levels (see \cref{tab:limits}) results in 3 rather stable latency levels.

\begin{figure}
    \centering
    \includegraphics[width=\linewidth]{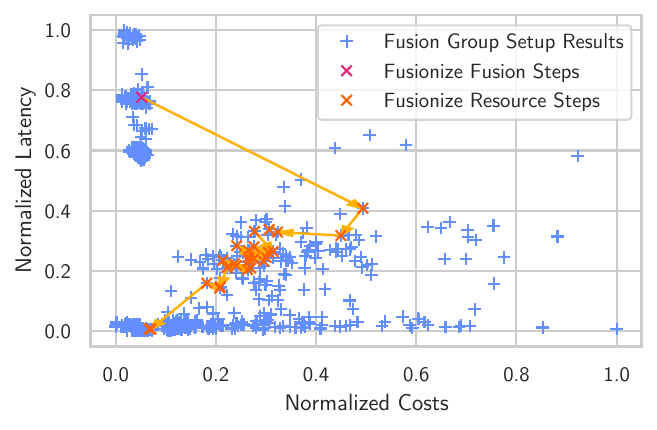}
    \caption{Benchmarking results for the \texttt{ASYNC} application shown as normalized latency versus normalized cost. 
    The highlighted optimization trajectory of Fusionize~\cite{schirmer2024fusionizepp} again coincides with the optimal fusion group configurations identified by our emulation-based FaaS benchmarking platform, despite the added complexity of asynchronous execution semantics. 
    This confirms that the offline emulation environment reliably captures the cost--latency trade-offs of asynchronous workloads and exposes further optimization potential in the near-origin region of the design space.
}
    \label{fig:async}
\end{figure}

The benchmarking results for the \texttt{ASYNC} application with its 768 different fusion group setups are displayed in \cref{fig:async}.
The different fusion groups are scattered, with distinct clusters forming across different cost levels.
The three distinct clusters with high latency values and low costs are caused by subpar resource allocations to the different function instances and fusion groups.

\subsection{Feasibility Study}
To validate the trends observed in our local emulation, we perform a targeted set of verification experiments on AWS Lambda.
Due to the cost and time constraints that motivate this work, exhaustive benchmarking of all possible configurations on a live platform is infeasible.
To ensure a comprehensive characterization of the system, we select a representative ensemble of configurations that spans the entire predicted performance spectrum, rather than focusing exclusively on top-tier results.

The mean optimization on AWS Lambda (51.42\%) closely approximates the local baseline (64.50\%).
While the primary intent of this evaluation was to demonstrate relative trends, these absolute values remain notably competitive despite the inherent network overhead of a cloud environment.
Consequently, the results not only confirm the expected performance dynamics but also demonstrate an additional layer of optimization: the slight delta in latency reduction is effectively balanced by superior cost efficiency, maintaining the high consistency of our overall efficiency model.

Crucially, the configurations identified as ``optimal'' by \name{} consistently ranked within the top 5\% of performing sets on AWS Lambda. 
This confirms that the platform-imposed overheads act as a near-constant penalty across all configurations and do not invalidate the underlying decision logic for fusion.
Specifically, the Lambda experiments confirm our primary qualitative findings: fusing synchronous calls consistently improves performance, while maintaining separation for asynchronous calls remains superior under instance-based pricing models.

In summary, while our emulation provides a more optimistic view of absolute gains, it serves as a reliable proxy for \textit{rank-order optimization}.
\name{} accurately captures the relative ordering of fusion setups, providing actionable guidance for real-world deployments without the prohibitive costs of live-platform benchmarking.

\subsection{Findings}
\label{sub:dominantfs}
Our results align closely with current state-of-the-art optimization frameworks~\cite{schirmer2024fusionizepp}.
We achieve an 89\% latency reduction (compared to 82\% in prior work) and a 40\% cost reduction (vs. 42\%) relative to a baseline of deployed serverless applications without resource adjustments or fusion groups.
Overall latency improvements remain within a 2 percentage point margin.
Cost calculations are performed using the official pricing models of AWS Lambda (traditional) and Google Cloud Run Functions (instance-based), and validated against the respective cost calculators~\cite{gcloud_calc,aws_calc}.  
To further validate our findings, we conducted a targeted feasibility study on AWS Lambda, focusing on configurations predicted by \name{} to be optimal or near-optimal.
While the absolute gains in the cloud are somewhat lower due to network and platform overheads, the qualitative trends are consistent: the optimal configurations identified by \name{} rank within the top 5\% on Lambda, confirming that emulation-based rank-order behavior translates to real-world deployments.

\begin{figure}
    \centering
    \includegraphics[width=\linewidth]{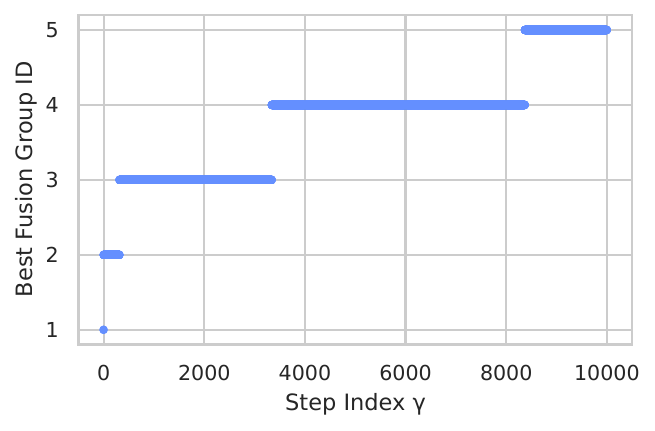}
    \caption{Optimal fusion configurations for the \texttt{TREE} application across the cost-latency trade-off index $\gamma$, utilizing the traditional pricing model. Only five distinct configurations emerge as optimal across the entire parameter sweep. Their respective coverage of the solution space is: ID 1 (0.02\%), ID 2 (3.14\%), ID 3 (30.26\%), ID 4 (50.29\%), and ID 5 (16.3\%). These results demonstrate that a small subset of configurations is sufficient to satisfy the full spectrum of latency and cost requirements.} 
    \label{fig:cost_weight_dominating}
\end{figure}

\paragraph{LINEAR Application}
Using the traditional pricing model, fusing all functions (\texttt{ABCDE}) is optimal for every $\alpha$.
This also holds for instance-based pricing.
\emph{Insight:} Preventing double billing dominates the optimization, making the fusion decision independent of cost-latency weighting.

\begin{figure}
    \centering
    \begin{subfigure}{.7\linewidth}
        \centering
        \includegraphics[width=.6\linewidth]{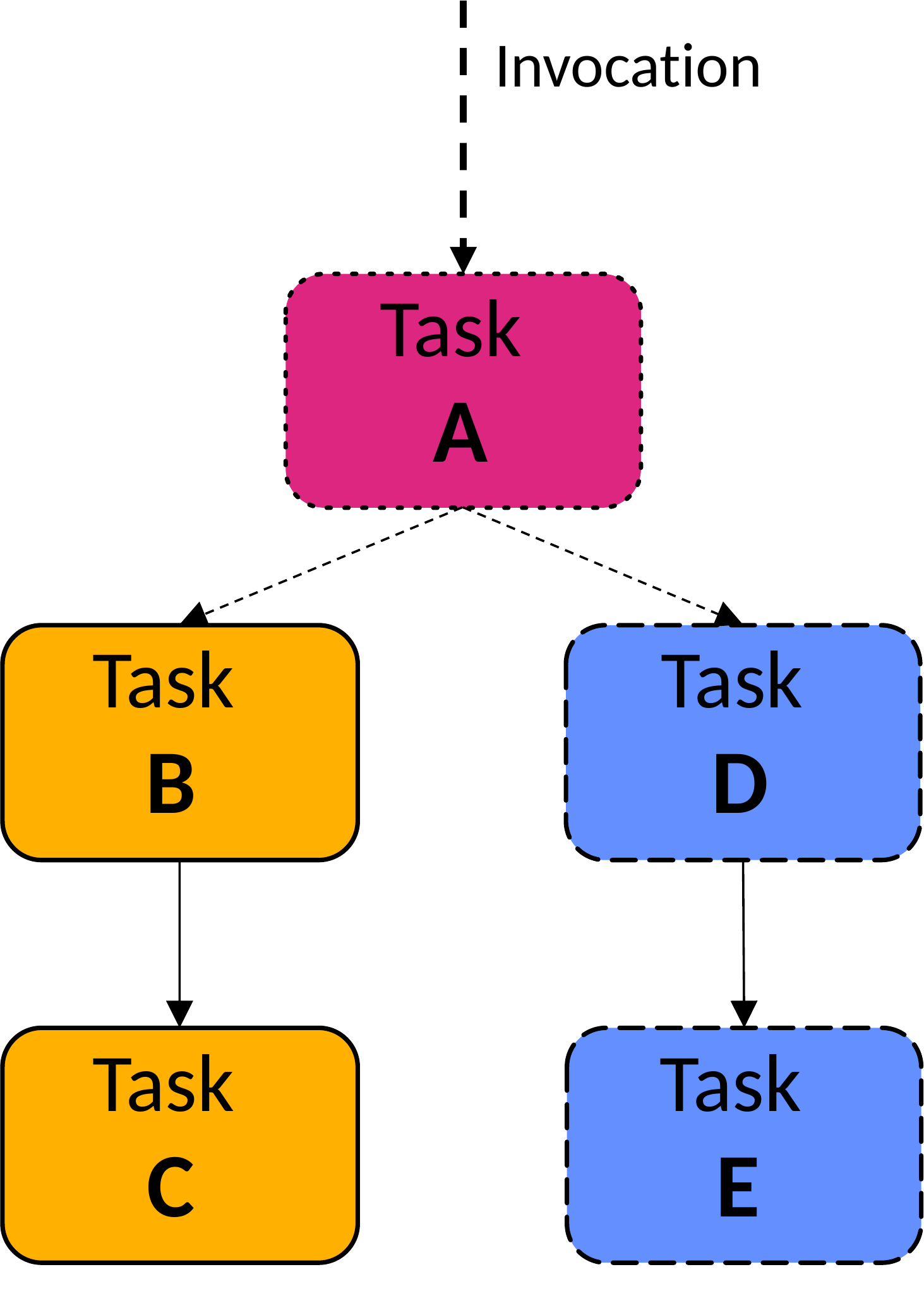}
        \caption{Optimal fusion configuration for the \texttt{PARALLEL-LINEAR} application.}
        \label{fig:parallel-linear:opt}
    \end{subfigure}
    \begin{subfigure}{.7\linewidth}
        \centering
        \includegraphics[width=\linewidth]{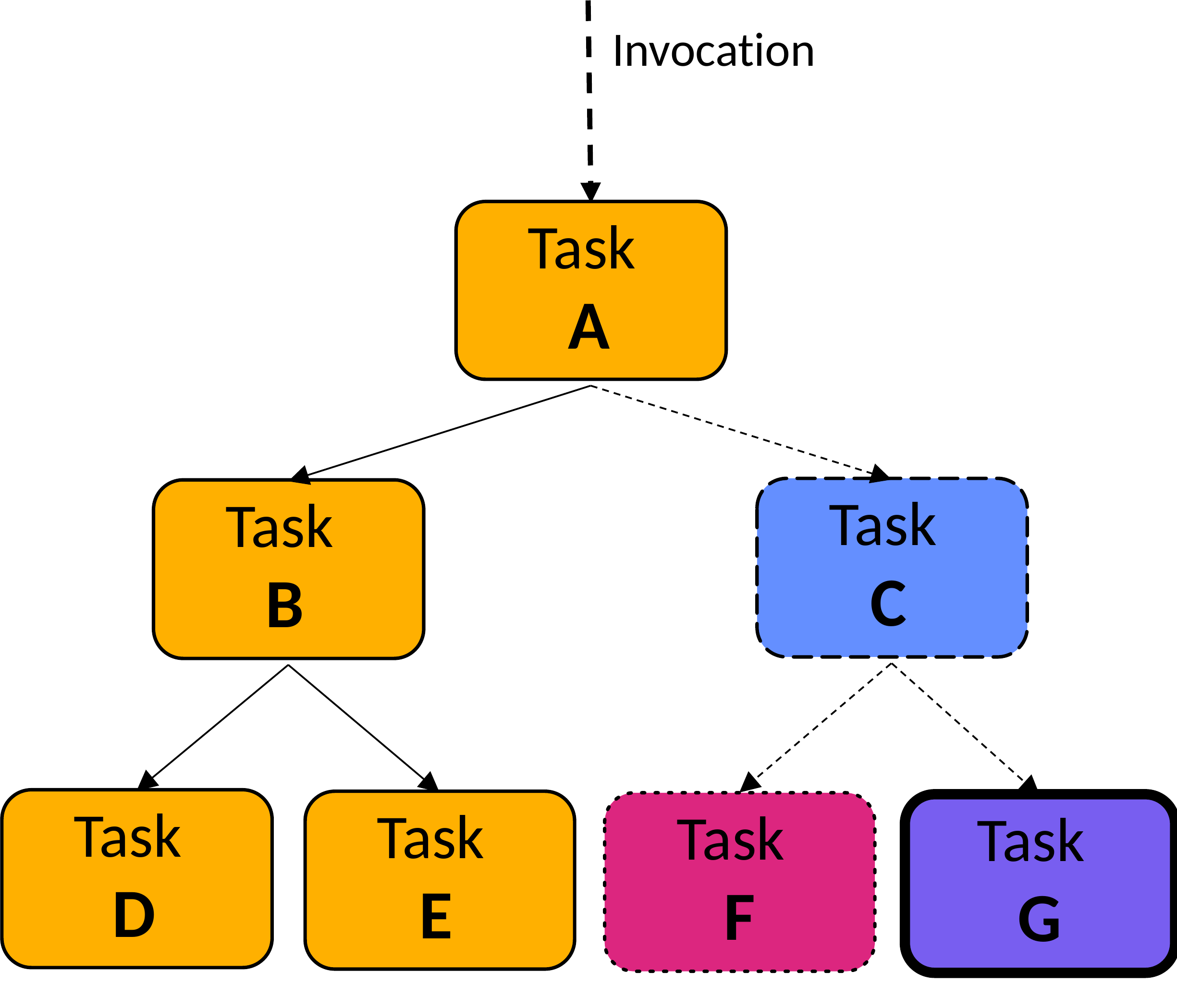}
        \caption{Optimal fusion configuration for the \texttt{TREE} application.}
        \label{fig:tree:opt}
    \end{subfigure}
    \caption{The optimal fusion group configurations for the \texttt{PARALLEL-LINEAR} (a) and \texttt{TREE} (b) applications for any latency or cost weights when using an instance-based pricing model. Here the best configuration for the \texttt{PARALLEL-LINEAR} application is to create the 3 fusion groups: \texttt{A}, \texttt{BC}, and \texttt{DE}.
    The best configuration for the \texttt{TREE} application is to build the 4 fusion groups: \texttt{ABDE}, \texttt{C}, \texttt{F}, and \texttt{G}.}
    \label{fig:tree_opt}
\end{figure}

\paragraph{PARALLEL-LINEAR Application}
Three configurations emerge as optimal under traditional pricing: \texttt{A,BC,DE}, \texttt{ADE,BC}, and \texttt{ABC,DE}.
\texttt{A,BC,DE} (all synchronous tasks fused) is optimal for 15.86\% of $\alpha$, the remaining 84.14\% split between the other two configurations.
For instance-based pricing, one configuration (all synchronous calls fused) dominates for any $\alpha$ (see \cref{fig:parallel-linear:opt}).
\emph{Insight:} Instance-based pricing stabilizes optimal fusion choices across weightings.

\paragraph{TREE Application}
Traditional pricing shows three main configurations dominating 96.84\% of cases:
\texttt{ABDE,CFG} (50.28\%), \texttt{ABDE,CF,G} (30.26\%), and \texttt{ABDE,C,F,G} (16.30\%), while minor outliers appear at extreme $\alpha$ values.
Instance-based pricing: \texttt{ABDE,C,F,G} is optimal for all $\alpha$ (see \cref{fig:tree:opt}).
\emph{Insight:} Synchronous fusion is consistently optimal; asynchronous fusion varies based on cost-latency trade-offs.

\paragraph{ASYNC Application}
Traditional pricing: keeping all asynchronous functions separate (\texttt{A,B,C,D,E}) is optimal for only 2.56\% of $\alpha$, with the majority fused in \texttt{ABD,C,E} (60.97\%) or \texttt{A,BC,DE} (36.47\%).
Instance-based pricing: separation (\texttt{A,B,C,D,E}) becomes optimal in 57.79\% of cases, with remaining $\alpha$ split between partial fusion configurations.
\emph{Insight:} Pricing model strongly influences optimal fusion for asynchronous functions.

\paragraph{Overall Patterns}
\begin{itemize}
    \item Synchronous chains should always be fused to reduce latency and double billing.  
    \item Asynchronous branches can remain separate unless instance-based pricing dominates.  
    \item Optimal fusion decisions are largely robust across $\alpha$ values and pricing models, especially for synchronous tasks.  
    \item Emulation rankings reliably reflect rank-order optimization, providing actionable guidance for cloud deployments without exhaustive benchmarking.
\end{itemize}
\section{Implications and Discussion}
\label{sec:discussion}

Our experiments with \name{} show that only a few fusion group configurations are actually meaningful, and the framework enables detailed cost--latency analysis to identify the best configuration for each application.
By systematically exploring the design space, \name{} allows us to derive heuristics that provide consistently good results across workloads, at costs far lower than full cloud benchmarking.
These findings can also be validated in real cloud environments, confirming that \name{} serves as a middle-ground solution: more accurate than simulations or heuristics, but cheaper and faster than running exhaustive experiments directly in the cloud.

\paragraph{Impact of pricing models} Our experiments highlight that the choice of pricing model substantially affects optimal fusion group configurations.
Asynchronous function calls interact differently with pricing: traditional per-request pricing penalizes non-fused functions, whereas instance-based runtime billing favors keeping asynchronous calls separate.

For instance-based pricing, we observe in our experiments a single dominant fusion group configuration across scoring methods (see \cref{sub:dominantfs}).
One setup consistently performs well for any weighting of cost and latency (see \cref{fig:tree_opt}).
Traditional pricing, in contrast, produces multiple optimal configurations depending on the cost--latency ratio ($\alpha$) (see \cref{fig:cost_weight_dominating}).
This demonstrates that asynchronous function optimization is sensitive to the pricing model, particularly when cost is not contributing with a factor of 0 (so $\alpha \neq 1$).

\paragraph{Heuristic rules}
Across our experiments, simple heuristics approximate the dominant configurations under instance-based pricing and non-extreme latency--cost ratios:
\begin{enumerate}
\item Always fuse synchronous calls, independent of the pricing model, to eliminate \emph{double billing}.
\item Keep asynchronous calls separate under instance-based pricing to exploit independent scaling effects.
\end{enumerate}

These empirically derived rules are consistent with those previously reported in prior work~\cite{elgamalCostlessOptimizingCost2018,tirkeyNovelFunctionFusion2023,schirmer2024fusionizepp,liEnhancingEffectiveBidirectional2024}.
They therefore provide a concise synthesis of existing insights while clarifying the pricing regimes and workload characteristics under which such heuristics remain applicable.

\paragraph{Critical reflection and verification} While \name{} produces clear trends in emulation, verification on AWS Lambda shows that no single fusion group consistently outperforms all others.
Instead, Lambda exhibits a cluster of similarly performing configurations.
Absolute improvements in \name{} are slightly optimistic, but the rank-ordering of configurations is preserved: our predicted fusion groups are consistently within the top 5\% on Lambda.
This confirms that \name{} supports rank-order optimization, which is the primary objective, and also produces absolute performance values that align closely with reality.

\paragraph{Comparison to related work} Unlike existing heuristics or simulations, \name{} executes locally and emulates cloud behavior, producing results that are more reliable than heuristic approximations but less costly than full cloud evaluation.
When compared to Fusionize, the most advanced prior work, \name{} demonstrates similar trends, validating the heuristics derived from the literature while providing additional experiment-based confirmation.

An alternative to emulation would be to fit a performance model per function and then simulate their composition, which is cheaper once the models exist.
We chose emulation for two reasons.
First, such models must themselves be calibrated from measurements, and the composition effects that matter here --- runtime contention inside a fused group, the cost of crossing a group boundary, cold-start behavior --- are properties of the composition rather than of the individual functions, so they are what a per-function model is least able to supply.
Second, emulation remains valid when the system beneath the application changes, for example under different resource limits, container runtimes, or host hardware, whereas a fitted model must be recalibrated for each such change.
We therefore see the two as complementary: \name{} is well suited to generating the data from which such models could be built, and we regard that as a promising direction rather than a competing one.

\paragraph{Applying \name{} to an application}
The intended workflow is not to match an application against our four topologies and reuse the corresponding result.
Users supply their own application as a DAG of tasks, and \name{} enumerates and measures its configuration space directly; the topologies in this paper are the instrument by which we derived the heuristics, not a catalogue to look up.
For applications small enough for exhaustive search, the ranking for the user's chosen $\alpha$ is the answer, and we recommend validating the top-$k$ configurations in the cloud before committing.
For larger applications, where exhaustive search is currently impractical, our heuristics --- fuse synchronous chains, keep asynchronous branches separate under instance-based pricing --- provide a starting configuration that can then be refined locally.
\paragraph{Threats to validity}
\textbf{Construct validity.} \name{} does not observe billed invoices; it derives cost by applying published provider pricing models to measured execution and communication metrics.
Cost is therefore a modeled quantity, which we mitigate by validating our calculations against the official provider calculators~\cite{gcloud_calc,aws_calc}.

\textbf{Internal validity.} Workloads are work-bounded and deterministic and all input parameters are fixed (\cref{sec:approach:applications}), so observed differences are attributable to fusion and resource allocation rather than to workload variance.
Evaluation under load is nevertheless limited to a bounded number of requests, which may affect observations under high concurrency, and rare-event behavior is underrepresented --- notably cascading cold starts, for which fusion is particularly consequential~\cite{paper_bermbach2020_faas_coldstarts,Daw_Bellur_Kulkarni_2020}.

\textbf{External validity.} Our benchmark applications are CPU-bound and synthetic: they contain no external service calls, no I/O, and no workload variability, and \name{} does not model connections beyond the FaaS application itself.
Local emulation likewise cannot reproduce heterogeneous provider hardware or multi-tenant interference.
In our Lambda verification these effects behaved as a near-uniform penalty that preserved rank order rather than reordering configurations, but applications whose latency is dominated by I/O or by external services may fuse differently, and we do not claim the derived heuristics transfer to them unchanged.

\textbf{Conclusion validity.} We state results as rank-order rather than absolute claims.
\name{} is optimistic about the magnitude of improvement relative to Lambda, which is why we report that predicted configurations fall within the top 5\% on Lambda rather than that they are globally optimal.

\textbf{Scalability.} The configuration space grows combinatorially with the number of tasks (\cref{sec:approach}), and measurement time grows with it: with three resource levels, exhaustive evaluation scales from 8 hours for 5 functions to 176 hours for 7.
Exhaustive search is thus practical for the application sizes studied here but not for large graphs.
Three mitigations are compatible with the present design and are the natural next step: pruning configurations that are dominated on both axes before measuring them, parallelizing independent configurations across hosts, and reusing measurements of fusion groups that recur unchanged across configurations, since a group's behavior depends only on its own composition and resource limit.
We have not implemented these, so the reported figures characterize the unoptimized prototype.

\paragraph{Future directions} Despite these limitations, \name{} provides actionable insights for optimizing FaaS applications:

\begin{itemize}
\item Investigating optimal resource allocation for given fusion groups.
\item Evaluating the role of asynchronous communication in real-world deployments.
\item Extending verification to larger, more diverse serverless workloads beyond current benchmarks.
\end{itemize}

Overall, while \name{} overestimates absolute improvements compared to AWS Lambda, the qualitative trends—especially the impact of pricing models and the effectiveness of fusing synchronous calls—hold true.
\name{} therefore supports both heuristic analysis at acceptable cost and rank-order optimization, serving as a practical middle-ground solution for FaaS application tuning.
\section{Conclusion}
\label{sec:conclusion}

In this paper, we presented \name{}, a local FaaS emulation framework designed to navigate the combinatorial complexity of function fusion.
By integrating provider-specific cost models with local execution, \name{} enabled the systematic exploration of configuration spaces that were previously too expensive to benchmark in the cloud. 

Our evaluation of multiple serverless applications demonstrated that only a small fraction of fusion configurations approached optimality, and that these near-optimal configurations were strongly influenced by the underlying pricing models.
Furthermore, validation on AWS Lambda confirmed that \name{} successfully captured qualitative performance trends, consistently ranking high-performing configurations accurately.
These findings underscored the feasibility of using high-fidelity emulation as a proxy for exhaustive cloud experimentation.
Ultimately, \name{} provides a cost-effective foundation for future research into automated, performance-aware FaaS deployment strategies and resource optimization in serverless environments.

\balance
\newpage
\bibliographystyle{IEEEtran}
\bibliography{bibliography.bib}

@article{Jonas_2019_ViewOn,
    title={Cloud Programming Simplified: A Berkeley View on Serverless Computing},
    note={arXiv: 1902.03383},
    journal={arXiv:1902.03383 [cs]},
    author={Jonas, Eric and Schleier-Smith, Johann and Sreekanti, Vikram and Tsai, Chia-Che and Khandelwal, Anurag and Pu, Qifan and Shankar, Vaishaal and Carreira, Joao and Krauth, Karl and Yadwadkar, Neeraja and Gonzalez, Joseph E. and Popa, Raluca Ada and Stoica, Ion and Patterson, David A.},
    year={2019}
}

@inproceedings{Brooker_2023_OnDemand,
    title={On-demand Container Loading in {AWS} Lambda},
    ISBN={978-1-939133-35-9},
    url={https://www.usenix.org/conference/atc23/presentation/brooker},
    author={Brooker, Marc and Danilov, Mike and Greenwood, Chris and Piwonka, Phil},
    year={2023},
    pages={315–328},
    language={en}
}

@article{Brooker_2021_Uniqueness,
    title={Restoring Uniqueness in MicroVM Snapshots},
    url={http://arxiv.org/abs/2102.12892},
    note={arXiv:2102.12892 [cs]},
    number={arXiv:2102.12892},
    publisher={arXiv},
    author={Brooker, Marc and Catangiu, Adrian Costin and Danilov, Mike and Graf, Alexander and MacCarthaigh, Colm and Sandu, Andrei},
    year={2021},
    month=feb,
    language={en}
}

@inproceedings{Sahraei_2023_XFaaS,
    address={New York, NY, USA},
    series={SOSP ’23},
    title={XFaaS: Hyperscale and Low Cost Serverless Functions at Meta},
    ISBN={979-8-4007-0229-7},
    url={https://doi.org/10.1145/3600006.3613155},
    DOI={10.1145/3600006.3613155},
    booktitle={Proceedings of the 29th Symposium on Operating Systems Principles},
    publisher={Association for Computing Machinery},
    author={Sahraei, Alireza and Demetriou, Soteris and Sobhgol, Amirali and Zhang, Haoran and Nagaraja, Abhigna and Pathak, Neeraj and Joshi, Girish and Souza, Carla and Huang, Bo and Cook, Wyatt and Golovei, Andrii and Venkat, Pradeep and Mcfague, Andrew and Skarlatos, Dimitrios and Patel, Vipul and Thind, Ravinder and Gonzalez, Ernesto and Jin, Yun and Tang, Chunqiang},
    year={2023},
    month=oct,
    pages={231–246},
    collection={SOSP ’23}
}

@article{schirmer2024fusionizepp,
    author = "Schirmer, Trever and Scheuner, Joel and Pfandzelter, Tobias and Bermbach, David",
    title = "FUSIONIZE++: Improving Serverless Application Performance Using Dynamic Task Inlining and Infrastructure Optimization",
    journal = "IEEE Transactions on Cloud Computing",
    volume = MISSING,
    number = MISSING,
    month = aug,
    year = 2024,
    issn = "2168-7161",
    pages = MISSING,
    articleno = MISSING,
    numpages = 16,
    url = "https://doi.org/10.1109/TCC.2024.3451108",
    doi = "10.1109/TCC.2024.3451108",
    acmid = MISSING,
    publisher = "IEEE",
    address = "New York, NY, USA",
    issue_date = MISSING,
    eprint = MISSING
}

@inproceedings{paper_bermbach2020_faas_coldstarts,
    author = "Bermbach, David and Karakaya, Ahmet-Serdar and Buchholz, Simon",
    title = "Using Application Knowledge to Reduce Cold Starts in FaaS Services",
    booktitle = "Proceedings of the 35th ACM Symposium on Applied Computing",
    pages = "134--143",
    month = mar,
    year = 2020,
    acmid = MISSING,
    publisher = "ACM",
    address = "New York, NY, USA",
    series = "SAC '20",
    location = "Brno, Czech Republic",
    numpages = MISSING,
    url = "https://doi.org/10.1145/3341105.3373909",
    doi = "10.1145/3341105.3373909"
}

@inproceedings{paper_pfandzelter2020_tinyfaas,
    author = "Pfandzelter, Tobias and Bermbach, David",
    title = "{tinyFaaS}: A Lightweight FaaS Platform for Edge Environments",
    booktitle = "Proceedings of the Second {IEEE} International Conference on Fog Computing",
    pages = "17--24",
    month = apr,
    year = 2020,
    acmid = MISSING,
    publisher = "IEEE",
    address = "New York, NY, USA",
    series = "ICFC 2020",
    location = "Sydney, NSW, Australia",
    numpages = MISSING,
    url = "https://doi.org/10.1109/ICFC49376.2020.00011",
    doi = "10.1109/ICFC49376.2020.00011"
}

@inproceedings{paper_schirmer2022_fusionize,
    author = "Schirmer, Trever and Scheuner, Joel and Pfandzelter, Tobias and Bermbach, David",
    title = "Fusionize: Improving Serverless Application Performance through Feedback-Driven Function Fusion",
    booktitle = "Proceedings of the 10th IEEE International Conference on Cloud Engineering",
    pages = "85--95",
    month = sep,
    year = 2022,
    acmid = MISSING,
    publisher = "IEEE",
    address = "New York, NY, USA",
    series = "IC2E 2022",
    location = "Asilomar, CA, USA",
    numpages = MISSING,
    url = "https://doi.org/10.1109/IC2E55432.2022.00017",
    doi = "10.1109/IC2E55432.2022.00017"
}

@inproceedings{schirmer2023nightshift,
    author = "Schirmer, Trever and Japke, Nils and Greten, Sofia and Pfandzelter, Tobias and Bermbach, David",
    title = "The Night Shift: Understanding Performance Variability of Cloud Serverless Platforms",
    booktitle = "Proceedings of the 1st Workshop on SErverless Systems, Applications and MEthodologies",
    month = may,
    year = 2023,
    acmid = MISSING,
    publisher = "ACM",
    address = "New York, NY, USA",
    series = "SESAME '23",
    location = "Rome, Italy",
    numpages = 6,
    url = "https://doi.org/10.1145/3592533.3592808",
    doi = "10.1145/3592533.3592808"
}

@inproceedings{schirmer2023profaastinate,
    author = "Schirmer, Trever and Carl, Valentin and Pfandzelter, Tobias and Bermbach, David",
    title = "ProFaaStinate: Delaying Serverless Function Calls to Optimize Platform Performance",
    booktitle = "Proceedings of the 9th International Workshop on Serverless Computing",
    pages = "1--6",
    month = dec,
    year = 2023,
    acmid = MISSING,
    publisher = "ACM",
    address = "New York, NY, USA",
    series = "WoSC '23",
    location = "Bologna, Italy",
    numpages = 6,
    url = "https://doi.org/10.1145/3631295.3631393",
    doi = "10.1145/3631295.3631393"
}

@online{awsLambda,
  title = {AWS Lambda Pricing},
  url = {https://aws.amazon.com/lambda/pricing/},
  lastaccessed = {2026-02-13},
  year = {2026},
  langid = {american},
  organization = {Amazon Web Services, Inc.},
  note = {Accessed: Apr. 30, 2026},
}

@article{czentyeServerlessApplicationComposition2024,
  title = {Serverless Application Composition Leveraging Function Fusion: {{Theory}} and Algorithms},
  author = {Czentye, János and Sonkoly, Balázs},
  date = {2024},
  year = {2024},
  journal = {Future Generation Computer Systems},
  volume = {153},
  pages = {403--418},
  issn = {0167-739X},
  doi = {10.1016/j.future.2023.12.010},
  url = {https://www.sciencedirect.com/science/article/pii/S0167739X23004648}
}

@article{mahgoubWISEFUSEWorkloadCharacterization2022,
  title = {{{WISEFUSE}}: {{Workload Characterization}} and {{DAG Transformation}} for {{Serverless Workflows}}},
  shorttitle = {{{WISEFUSE}}},
  author = {Mahgoub, Ashraf and Yi, Edgardo Barsallo and Shankar, Karthick and Minocha, Eshaan and Elnikety, Sameh and Bagchi, Saurabh and Chaterji, Somali},
  date = {2022-05-26},
  year = {2022},
  journal = {Proceedings of the ACM on Measurement and Analysis of Computing Systems},
  shortjournal = {Proc. ACM Meas. Anal. Comput. Syst.},
  volume = {6},
  number = {2},
  pages = {1--28},
  issn = {2476-1249},
  doi = {10.1145/3530892},
  url = {https://dl.acm.org/doi/10.1145/3530892},
  urldate = {2024-12-16},
  langid = {english}
}

@inproceedings{sheshadriRightFusionEnablingQoS2024,
  title = {{{RightFusion}}: {{Enabling QoS}} Driven {{Function Fusion}} in {{Edge-Cloud FaaS}}},
  shorttitle = {{{RightFusion}}},
  booktitle = {2024 {{IEEE International Conference}} on {{Cloud Engineering}} ({{IC2E}})},
  author = {Sheshadri, K R and Lakshmi, J},
  date = {2024-09},
  year = {2024},
  pages = {160--167},
  issn = {2694-0825},
  doi = {10.1109/IC2E61754.2024.00025},
  url = {https://ieeexplore.ieee.org/document/10749776/?arnumber=10749776},
  urldate = {2024-12-16},
  eventtitle = {2024 {{IEEE International Conference}} on {{Cloud Engineering}} ({{IC2E}})}
}

@inproceedings{elgamalCostlessOptimizingCost2018,
  title = {Costless: {{Optimizing Cost}} of {{Serverless Computing}} through {{Function Fusion}} and {{Placement}}},
  shorttitle = {Costless},
  booktitle = {2018 {{IEEE}}/{{ACM Symposium}} on {{Edge Computing}} ({{SEC}})},
  author = {Elgamal, Tarek and Sandur, Atul and Nahrstedt, Klara and Agha, Gul},
  date = {2018-10},
  year = {2018},
  pages = {300--312},
  doi = {10.1109/SEC.2018.00029},
  url = {https://ieeexplore.ieee.org/document/8567674},
  urldate = {2024-08-08},
  eventtitle = {2018 {{IEEE}}/{{ACM Symposium}} on {{Edge Computing}} ({{SEC}})}
}

@inproceedings{akhtarCOSEConfiguringServerless2020a,
  title = {{{COSE}}: {{Configuring Serverless Functions}} Using {{Statistical Learning}}},
  shorttitle = {{{COSE}}},
  booktitle = {{{IEEE INFOCOM}} 2020 - {{IEEE Conference}} on {{Computer Communications}}},
  author = {Akhtar, Nabeel and Raza, Ali and Ishakian, Vatche and Matta, Ibrahim},
  date = {2020-07},
  year = {2020},
  pages = {129--138},
  issn = {2641-9874},
  doi = {10.1109/INFOCOM41043.2020.9155363},
  url = {https://ieeexplore.ieee.org/document/9155363/?arnumber=9155363},
  urldate = {2025-01-14},
  eventtitle = {{{IEEE INFOCOM}} 2020 - {{IEEE Conference}} on {{Computer Communications}}}
}

@inproceedings{cordinglyFunctionMemoryOptimization2022,
  title = {Function {{Memory Optimization}} for {{Heterogeneous Serverless Platforms}} with {{CPU Time Accounting}}},
  booktitle = {2022 {{IEEE International Conference}} on {{Cloud Engineering}} ({{IC2E}})},
  author = {Cordingly, Robert and Xu, Sonia and Lloyd, Wes},
  date = {2022-09},
  year = {2022},
  pages = {104--115},
  doi = {10.1109/IC2E55432.2022.00019},
  url = {https://ieeexplore.ieee.org/document/9946331},
  urldate = {2025-01-14},
  eventtitle = {2022 {{IEEE International Conference}} on {{Cloud Engineering}} ({{IC2E}})}
}

@inproceedings{baldiniServerlessTrilemmaFunction2017a,
  title = {The Serverless Trilemma: Function Composition for Serverless Computing},
  shorttitle = {The Serverless Trilemma},
  booktitle = {Proceedings of the 2017 {{ACM SIGPLAN International Symposium}} on {{New Ideas}}, {{New Paradigms}}, and {{Reflections}} on {{Programming}} and {{Software}}},
  author = {Baldini, Ioana and Cheng, Perry and Fink, Stephen J. and Mitchell, Nick and Muthusamy, Vinod and Rabbah, Rodric and Suter, Philippe and Tardieu, Olivier},
  date = {2017-10-25},
  year = {2017},
  series = {Onward! 2017},
  pages = {89--103},
  publisher = {Association for Computing Machinery},
  location = {New York, NY, USA},
  doi = {10.1145/3133850.3133855},
  url = {https://dl.acm.org/doi/10.1145/3133850.3133855},
  urldate = {2025-01-14},
  isbn = {978-1-4503-5530-8}
}

@inproceedings{liEnhancingEffectiveBidirectional2024,
  title = {Enhancing {{Effective Bidirectional Isolation}} for {{Function Fusion}} in {{Serverless Architectures}}},
  booktitle = {Proceedings of the 25th {{International Middleware Conference}}},
  author = {Li, Tianyu and Chen, Yingpeng and Yu, Donghui and Zhang, Yuanyuan and Lagaisse, Bert},
  date = {2024-12-02},
  year = {2024},
  pages = {1--7},
  publisher = {ACM},
  location = {Hong Kong Hong Kong},
  doi = {10.1145/3652892.3654778},
  url = {https://dl.acm.org/doi/10.1145/3652892.3654778},
  urldate = {2024-12-16},
  eventtitle = {{{MIDDLEWARE}} '24: 25th {{International Middleware Conference}}},
  isbn = {979-8-4007-0623-3},
  langid = {english}
}

@article{linFineGrainedPerformanceCost2023,
  title = {Fine-{{Grained Performance}} and {{Cost Modeling}} and {{Optimization}} for {{FaaS Applications}}},
  author = {Lin, Changyuan and Mahmoudi, Nima and Fan, Caixiang and Khazaei, Hamzeh},
  date = {2023-01},
  year = {2023},
  journaltitle = {IEEE Transactions on Parallel and Distributed Systems},
  volume = {34},
  number = {1},
  pages = {180--194},
  issn = {1558-2183},
  doi = {10.1109/TPDS.2022.3214783},
  url = {https://ieeexplore.ieee.org/document/9919386/?arnumber=9919386},
  urldate = {2024-08-08},
  eventtitle = {{{IEEE Transactions}} on {{Parallel}} and {{Distributed Systems}}}
}

@article{wenJointOptimizationParallelism2024,
  title = {Joint {{Optimization}} of {{Parallelism}} and {{Resource Configuration}} for {{Serverless Function Steps}}},
  author = {Wen, Zhaojie and Chen, Qiong and Niu, Yipei and Song, Zhen and Deng, Quanfeng and Liu, Fangming},
  date = {2024-04},
  year = {2024},
  journaltitle = {IEEE Transactions on Parallel and Distributed Systems},
  volume = {35},
  number = {4},
  pages = {560--576},
  issn = {1558-2183},
  doi = {10.1109/TPDS.2024.3365134},
  url = {https://ieeexplore.ieee.org/abstract/document/10433749},
  urldate = {2025-01-21},
  eventtitle = {{{IEEE Transactions}} on {{Parallel}} and {{Distributed Systems}}}
}

@inproceedings{rameshOptimalMappingWorkflows2024,
  title = {Optimal {{Mapping}} of {{Workflows Using Serverless Architecture}} in a {{Multi-Cloud Environment}}},
  booktitle = {2024 {{IEEE}} 21st {{International Conference}} on {{Software Architecture Companion}} ({{ICSA-C}})},
  author = {Ramesh, Manju and Phalak, Chetan and Chahal, Dheeraj and Singhal, Rekha},
  date = {2024-06},
  year = {2024},
  pages = {252--259},
  issn = {2768-4288},
  doi = {10.1109/ICSA-C63560.2024.00053},
  url = {https://ieeexplore.ieee.org/document/10628213/?arnumber=10628213},
  urldate = {2025-01-21},
  eventtitle = {2024 {{IEEE}} 21st {{International Conference}} on {{Software Architecture Companion}} ({{ICSA-C}})}
}

@unpublished{foxStatusServerlessComputing2017,
  title = {Status of {{Serverless Computing}} and {{Function-as-a-Service}} ({{FaaS}}) in {{Industry}} and {{Research}}},
  author = {Fox, Geoffrey C. and Ishakian, Vatche and Muthusamy, Vinod and Slominski, Aleksander},
  date = {2017},
  year = {2017},
  eprint = {1708.08028},
  eprinttype = {arXiv},
  eprintclass = {cs},
  doi = {10.13140/RG.2.2.15007.87206},
  url = {http://arxiv.org/abs/1708.08028},
  urldate = {2025-01-24},
}

@online{cvetkovicDirigentLightweightServerless2024,
  title = {Dirigent: {{Lightweight Serverless Orchestration}}},
  shorttitle = {Dirigent},
  author = {Cvetković, Lazar and Costa, François and Djokic, Mihajlo and Friedman, Michal and Klimovic, Ana},
  date = {2024-04-25},
  year = {2024},
  eprint = {2404.16393},
  eprinttype = {arXiv},
  eprintclass = {cs},
  url = {http://arxiv.org/abs/2404.16393},
  urldate = {2024-08-01},
  langid = {english},
  pubstate = {prepublished},
}

@inproceedings{klimovicPocketElasticEphemeral2018,
  title = {Pocket: {{Elastic Ephemeral Storage}} for {{Serverless Analytics}}},
  shorttitle = {Pocket},
  author = {Klimovic, Ana and Wang, Yawen and Stuedi, Patrick and Trivedi, Animesh and Pfefferle, Jonas and Kozyrakis, Christos},
  date = {2018},
  year = {2018},
  pages = {427--444},
  url = {https://www.usenix.org/conference/osdi18/presentation/klimovic},
  urldate = {2025-01-24},
  eventtitle = {13th {{USENIX Symposium}} on {{Operating Systems Design}} and {{Implementation}} ({{OSDI}} 18)},
  isbn = {978-1-939133-08-3},
  langid = {english},
}

@inproceedings{tirkeyNovelFunctionFusion2023,
  title = {A {{Novel Function Fusion Approach}} for {{Serverless Cold Start}}},
  booktitle = {2023 {{International Conference}} on {{Communication}}, {{Circuits}}, and {{Systems}} ({{IC3S}})},
  author = {Tirkey, Kanchan and Kumari, Anisha and Mohanty, Sagarika and Sahoo, Bibhudatta},
  date = {2023-05},
  year = {2023},
  pages = {1--5},
  doi = {10.1109/IC3S57698.2023.10169477},
  url = {https://ieeexplore.ieee.org/abstract/document/10169477},
  urldate = {2025-02-17},
  eventtitle = {2023 {{International Conference}} on {{Communication}}, {{Circuits}}, and {{Systems}} ({{IC3S}})},
}

@article{eismannwhy,
  author={Eismann, Simon and Scheuner, Joel and van Eyk, Erwin and Schwinger, Maximilian and Grohmann, Johannes and Herbst, Nikolas and Abad, Cristina L. and Iosup, Alexandru},
  journal={IEEE Software}, 
  title={Serverless Applications: Why, When, and How?}, 
  year={2021},
  volume={38},
  number={1},
  pages={32-39},
  doi={10.1109/MS.2020.3023302}}

@inproceedings{aslanpourServerlessEdgeComputing2021,
  title = {Serverless {{Edge Computing}}: {{Vision}} and {{Challenges}}},
  shorttitle = {Serverless {{Edge Computing}}},
  booktitle = {2021 {{Australasian Computer Science Week Multiconference}}},
  author = {Aslanpour, Mohammad S. and Toosi, Adel N. and Cicconetti, Claudio and Javadi, Bahman and Sbarski, Peter and Taibi, Davide and Assuncao, Marcos and Gill, Sukhpal Singh and Gaire, Raj and Dustdar, Schahram},
  date = {2021-02},
  year = {2021},
  pages = {1--10},
  publisher = {ACM},
  location = {Dunedin New Zealand},
  doi = {10.1145/3437378.3444367},
  url = {https://dl.acm.org/doi/10.1145/3437378.3444367},
  urldate = {2024-12-16},
  eventtitle = {{{ACSW}} '21: 2021 {{Australasian Computer Science Week Multiconference}}},
  isbn = {978-1-4503-8956-3},
  langid = {english},
}

@inproceedings{wangPeekingCurtainsServerless2018,
  title = {Peeking {{Behind}} the {{Curtains}} of {{Serverless Platforms}}},
  author = {Wang, Liang and Li, Mengyuan and Zhang, Yinqian and Ristenpart, Thomas and Swift, Michael},
  date = {2018},
  year = {2018},
  pages = {133--146},
  url = {https://www.usenix.org/conference/atc18/presentation/wang-liang},
  urldate = {2025-01-24},
  eventtitle = {2018 {{USENIX Annual Technical Conference}} ({{USENIX ATC}} 18)},
  isbn = {978-1-939133-01-4},
  langid = {english},
}

@inproceedings {216031,
author = {Edward Oakes and Leon Yang and Dennis Zhou and Kevin Houck and Tyler Harter and Andrea Arpaci-Dusseau and Remzi Arpaci-Dusseau},
title = {{SOCK}: Rapid Task Provisioning with {Serverless-Optimized} Containers},
booktitle = {2018 USENIX Annual Technical Conference (USENIX ATC 18)},
year = {2018},
isbn = {978-1-931971-44-7},
address = {Boston, MA},
pages = {57--70},
url = {https://www.usenix.org/conference/atc18/presentation/oakes},
publisher = {USENIX Association},
month = jul
}

@inproceedings{Daw_Bellur_Kulkarni_2020, address={New York, NY, USA}, series={Middleware ’20}, title={Xanadu: Mitigating cascading cold starts in serverless function chain deployments}, ISBN={978-1-4503-8153-6}, url={https://dl.acm.org/doi/10.1145/3423211.3425690}, DOI={10.1145/3423211.3425690}, abstractNote={Organization of tasks as workflows are an essential feature to expand the applicability of the serverless computing framework. Existing serverless platforms are either agnostic to function chains (workflows as a composition of functions) or rely on naive provisioning and management mechanisms of the serverless framework---an example is that they provision resources after the trigger to each function in a workflow arrives thereby forcing a setup latency for each function in the workflow. In this work, we focus on mitigating the cascading cold start problem--- the latency overheads in triggering a sequence of serverless functions according to a workflow specification. We first establish the nature and extent of the cascading effects in cold start situations across multiple commercial server platforms and cloud providers. Towards mitigating these cascading overheads, we design and develop several optimizations, that are built into our tool Xanadu. Xanadu offers multiple instantiation options based on the desired runtime isolation requirements and supports function chaining with or without explicit workflow specifications. Xanadu’s optimizations to address the cascading cold start problem are built on speculative and just-in-time provisioning of resources. Our evaluation of the Xanadu system reveals almost complete elimination of cascading cold starts at minimal cost overheads, outperforming the available state of the art platforms. For even relatively short workflows, Xanadu reduces platform overheads by almost 18x compared to Knative and 10x compared to Apache Openwhisk.}, booktitle={Proceedings of the 21st International Middleware Conference}, publisher={Association for Computing Machinery}, author={Daw, Nilanjan and Bellur, Umesh and Kulkarni, Purushottam}, year={2020}, month=dec, pages={356–370}, collection={Middleware ’20} }

@inproceedings{lin_demystifying_2026,
	location = {New York, {NY}, {USA}},
	title = {Demystifying Serverless Costs on Public Platforms: Bridging Billing, Architecture, and {OS} Scheduling},
	isbn = {979-8-4007-2212-7},
	url = {https://dl.acm.org/doi/10.1145/3767295.3769374},
	doi = {10.1145/3767295.3769374},
	series = {{EUROSYS} '26},
	shorttitle = {Demystifying Serverless Costs on Public Platforms},
	pages = {1964--1981},
	booktitle = {Proceedings of the 21st European Conference on Computer Systems},
	publisher = {Association for Computing Machinery},
	author = {Lin, Changyuan and Ma, Yuanzhi and Shahrad, Mohammad},
	urldate = {2026-04-29},
	date = {2026-04-26},
}

@inproceedings{mahgoub_orion_2022,
	title = {{ORION} and the Three Rights: Sizing, Bundling, and Prewarming for Serverless {DAGs}},
	isbn = {978-1-939133-28-1},
	url = {https://www.usenix.org/conference/osdi22/presentation/mahgoub},
	shorttitle = {{ORION} and the Three Rights},
	eventtitle = {16th {USENIX} Symposium on Operating Systems Design and Implementation ({OSDI} 22)},
	pages = {303--320},
	author = {Mahgoub, Ashraf and Yi, Edgardo Barsallo and Shankar, Karthick and Elnikety, Sameh and Chaterji, Somali and Bagchi, Saurabh},
	urldate = {2026-04-29},
	date = {2022},
	langid = {english},
}

@online{gcloud_calc,
    title = "{Google Cloud Pricing Calculator}",
    organization = "Google Cloud",
    url = "https://cloud.google.com/products/calculator",
    year = "2026",
    note = "Accessed: Apr. 30, 2026"
}

@online{aws_calc,
    title = "{AWS Pricing Calculator}",
    organization = "Amazon Web Services",
    url = "https://calculator.aws/#/",
    year = "2026",
    note = "Accessed: Apr. 30, 2026"
}

\end{document}